\documentclass{iau} 
\usepackage{graphicx}

\usepackage{xcolor}
\usepackage{amsmath}

\title[The Discovery of the Most Accelerated Binary Pulsar] %% give here short title %%
{The Discovery of the Most Accelerated Binary Pulsar}

\author[A.~D.~Cameron]   %% give here short author list %%
{Andrew~D.~Cameron$^1$\\ on behalf of the~HTRU~Collaboration and additional collaborators$^{2}$}

\affiliation{$^1$Max-Planck-Institut f{\"u}r Radioastronomie \\ Auf dem H{\"u}gel 69, D-53121 Bonn, Germany \\ email: {\tt acameron@mpifr-bonn.mpg.de} \\[\affilskip]
$^2$for full listing of collaborators on this work, see \cite[Cameron et al. (submitted)]{cam_J1757}}

\pubyear{2017}
\volume{337}  %% insert here IAU Symposium No.
\jname{Pulsar Astrophysics -­ The Next 50 Years}
\editors{P. Weltevrede, B.B.P. Perera, L. Levin Preston \& S. Sanidas, eds.}

\begin{document}

\maketitle

\begin{abstract}
Pulsars in relativistic binary systems have emerged as fantastic natural laboratories for testing theories of gravity, the most prominent example being the double pulsar, PSR~J0737$-$3039. The HTRU-South Low Latitude pulsar survey represents one of the most sensitive blind pulsar surveys taken of the southern Galactic plane to date, and its primary aim has been the discovery of new relativistic binary pulsars. Here we present our binary pulsar searching strategy and report on the survey's flagship discovery, PSR~J1757$-$1854. A 21.5-ms pulsar in a relativistic binary with an orbital period of 4.4 hours and an eccentricity of 0.61, this double neutron star (DNS) system is the most accelerated pulsar binary known, and probes a relativistic parameter space not yet explored by previous pulsar binaries.
%% add here a maximum of 10 keywords, to be taken form the file <Keywords.tex>
\end{abstract}

\firstsection % if your document starts with a section,
\section{Introduction}
As binary pulsars serve as some of the most useful laboratories in the exploration and testing of gravitational theories, the search for new relativistic binaries remains an active goal of pulsar astronomy. The unique properties of each newly discovered binary determine the nature of gravity tests to which it is best suited. For example, the millisecond pulsar-white dwarf binary PSR J1738$+$0333 strongly constrains scalar-tensor theories of gravity due to its large mass asymmetry (\cite[Freire et al. 2012]{fwe+12}), while the relativistic properties of the double pulsar (PSR~J0737$-$3039; \cite[Burgay et al. 2003]{bdp+03}; \cite[Lyne et al. 2004]{lyne04}) allow it to provide the strongest constraints on the quadrupole gravitational wave emission described by General Relativity (GR) (Kramer et al. in prep.). Ongoing searches for ever more exotic and extreme relativistic systems will ensure that binary pulsars continue to provide competitive and complementary tests of gravitational theories against current and future ground-based gravitational wave (GW) detectors (\cite[Shao et al. 2017]{shao2017}).

With this in mind, the primary goal of the the High Time Resolution Universe South Low Latitude pulsar survey (HTRU-S LowLat; \cite[Ng et al. 2015]{ncb15}), a component of the all-sky HTRU survey (\cite[Keith et al. 2010]{kjvs10}), has been the discovery of relativistic pulsars in tight binary systems. Conducted between 2008 and 2013 with the Parkes 64-m radio telescope, this survey covers the southern Galactic plane region ($-80^\circ<l<30^\circ$ and $\left|b\right|<3.5^\circ$) which is predicted to contain the highest number of relativistic binaries (\cite[Belczynski et al. 2002]{bkb02}). With each observation having a 72-min integration time, a nominal bandwidth of 400 MHz and a sampling time of $64\,\mu\text{s}$, this survey represents one of the most sensitive pulsar surveys of the Galactic plane region to date.

\section{The partially-coherent segmented acceleration search}

Any pulsar in a binary system will exhibit a Doppler modulation of its apparent rotational period due to its orbital motion. We employ a ``time-domain resampling'' acceleration search technique (see e.g., \cite[Middleditch \& Kristian 1984]{mk84}; \cite[Johnston \& Kulkarni]{jk91}) which assumes that this unknown orbital motion can be modeled as a constant acceleration. Prior demonstrations (see e.g., \cite[Johnston \& Kulkarni 1991]{jk91}; \cite[Ng et al. 2015]{ncb15}) have shown that for circular orbits, this assumption of constant acceleration holds best when $r_\text{b}={t_\text{int}}/{P_\text{b}}\leq0.1$, where $P_\text{b}$ is the orbital period and $t_\text{int}$ is the integration time of the observation (eccentricity will modify this optimal $r_\text{b}$ depending on the orbital phase at which the pulsar is observed). Assuming circularity, this implies that a 72-min HTRU-S LowLat observation is sensitive to $P_\text{b}\gtrsim12\,\text{hr}$.

In order to search for tighter binary systems, our ``partially-coherent segmented acceleration search'' (based on work by \cite[Eatough et al. 2013]{ekl+13}) progressively halves each observation into smaller segments before searching each independently with increasingly-large ranges of acceleration. With each halving of $t_\text{int}$, the minimum $P_\text{b}$ to which that segment is sensitive is also halved, but at the cost of a $\sqrt{2}$ loss in flux sensitivity. By searching at multiple values of $t_\text{int}$, our pipeline aims to strike an optimal balance between these two considerations. Table \ref{tab: segmented search} describes the basic structure and parameters of the segmented search, with full details available in \cite[Ng et al. (2015)]{ncb15}.

\begin{table}[t]
\begin{center}
\caption{A summary of the 15 individual segments searched as part of the partially-coherent segmented acceleration search. Each segment group spans the entire observation without overlap.}\label{tab: segmented search}
\scriptsize{
\begin{tabular}{|c|c|c|c|c|}
 \hline Segment & Amount. & $t_\text{int}$ (min) & Min. $P_\text{b}$ (hours) & $\left|a_\text{max}\right|$ ($\text{m\,s}^{-2}$) \\ \hline
 Full & 1 & 72 & 12 & 1 \\
 Half & 2 & 36 & 6 & 200 \\
 Quarter & 4 & 18 & 3 & 500 \\
 Eighth & 8 & 9 & 1.5 & 1200 \\ \hline 
\end{tabular}
}
\end{center}
\end{table}

\section{Discovery and confirmation of PSR~J1757$-$1854}\label{sec: confirmation}

PSR~J1757$-$1854, first identified in an observation recorded on MJD~56029, is a 21.5-ms pulsar in a 4.4-hr orbit around a neutron star (NS) companion, with an orbital eccentricity of $e\simeq0.61$. The segmented acceleration search initially detected the pulsar only in the second 36-min half-segment of the observation at an acceleration of approximately $-32\,\text{m\,s}^{-2}$ and a signal to noise ratio (S/N) of 13.3. Folding the observation at this acceleration (equivalent to a first period derivative $\dot{P}$) indicated a non-constant acceleration (i.e., jerk) across the 72-min span. This required the fitting of multiple period derivatives in order to sufficiently model the orbital motion and derive a maximal S/N of 21.4. 

As the detected acceleration of $-32\,\text{m\,s}^{-2}$ is larger than the $\left|a_\text{max}\right|=1\,\text{m\,s}^{-2}$ of the full-segment search, the full observation was re-searched with an $\left|a_\text{max}\right|=50\,\text{m\,s}^{-2}$ in order to verify role played by the segmented acceleration search in detecting PSR~J1757$-$1854. As a consequence of the jerk present in the observation, this new search returned a significantly reduced S/N of only 10.6, indicating the the segmentation aspect of our search contributed significantly to the discovery of PSR~J1757$-$1854.

\begin{figure}[t]
\begin{center}
 \includegraphics[height=0.9\textwidth, angle=270]{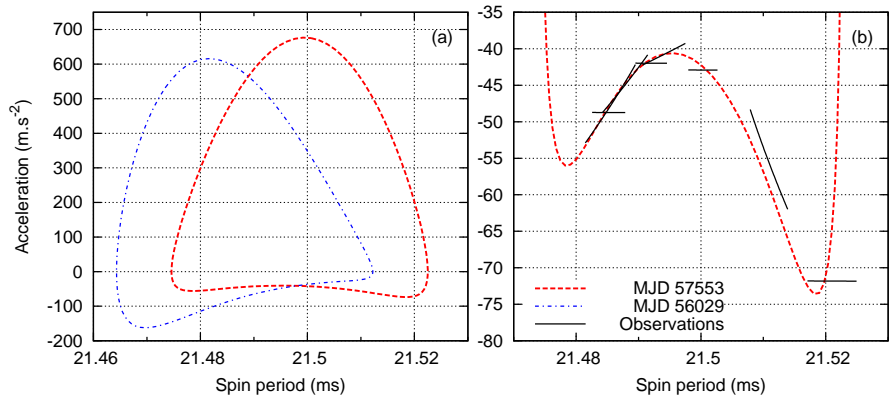} 
 \caption{Plot (a) shows the changing apparent line-of-sight spin period and acceleration caused by the orbit of PSR~J1757$-$1854 at both the epoch of discovery (MJD~56029) and the epoch at which the orbit was solved (MJD~57553). The change in shape is caused by the pulsar's high rate of periastron advance ($\dot{\omega}$). Plot (b) shows the observational data used to initially fit and constrain the orbit, as per the description in Section \ref{sec: confirmation}.}\label{fig: p-accel}
\end{center}
\end{figure}

Observations designed to first confirm and then map the orbit of the pulsar were conducted by the Parkes, Effelsberg and Jodrell Bank telescopes between MJD~$57405-57554$. The change in the apparent spin period and acceleration during each observation was modeled by a series of period derivatives so as to produce a Taylor series expansion for each parameter. The resulting polynomials were plotted on a period-acceleration diagram (see Figure \ref{fig: p-accel}). Following a method similar to that of \cite[Freire et al. (2001)]{fkl01}, these constraints were then used to fit an orbital solution.

During the confirmation campaign, it was found that the pulsar could not be detected by standard acceleration search techniques in multiple observations. Following the development of an orbital solution, it was determined that these observations took place during the orbit's periastron phase, an approximately 50 to 60-min interval where the pulsar's line-of-sight acceleration climbed from the range of negative values seen in Figure \ref{fig: p-accel}b to reach as high as $\sim+680\,\text{m\,s}^{-2}$ before returning\footnote{PSR~J1757$-$1854 reaches a maximum absolute acceleration of $\sim684\,\text{m\,s}^{-2}$, the highest of any known binary pulsar system.}. Naturally, this rendered the assumption of a constant acceleration inapplicable, causing standard search techniques to fail. Later ephemeris folds which incorporated the orbital solution were successful in recovering the signal of the pulsar in all observations with prior non-detections.

\section{Relativistic properties and future prospects}

Following a 1.6-year timing campaign with the Parkes, Jodrell Bank, Effelsberg and Green Bank telescopes, a full timing solution of PSR~J1757$-$1854, with measurements of five post-Keplerian (PK) parameters, has been determined. This solution, along with further discussion regarding the properties and future prospects of PSR~J1757$-$1854, is presented in a separate publication (\cite[Cameron et~al. submitted]{cam_J1757}). Under the assumption of GR, the PK parameters allow the total system mass $M=2.73295(9)\,\text{M}_\odot$ and the separate masses of the pulsar ($m_\text{p}=1.3384(9)\,\text{M}_\odot$) and its companion ($m_\text{c}=1.3946(9)\,\text{M}_\odot$) to be determined. These masses, combined with an eccentricity of $e=0.6058142(10)$ and an implied second supernova, clearly indicate that the system is a DNS. However, all searches for pulsations from the companion neutron star have to date been negative.

The parameters of PSR~J1757$-$1854 are such that it probes a new relativistic parameter space and breaks multiple records, including the closest binary separation ($0.749\,\text{R}_\odot$) and the highest relative velocity at periastron ($1060\,\text{km\,s}^{-1}$). It also possesses the highest recorded values of $\dot{P}_\text{b}=-5.3(2)\times10^{-12}$ and $\dot{P}_\text{b}/P_\text{b}=-3.33\times10^{-16}\,\text{s}^{-1}$ due to GW damping. PSR~J1757$-$1854 is also expected to allow for a measurement of the sixth PK parameter $\delta_\theta$ and constraint on the presence of Lense-Thirring precession to within $3\sigma$ in as few as 7-9 years. The measurability of $\delta_\theta$  is due to both a high $\dot{\omega}=10.3651(2)$ and high eccentricity (\cite[Damour \& Deruelle 1986]{dd86}), while a future measurement of Lense-Thirring precession (expected within a similar timeframe) comes as a result of the strong likelihood of a misalignment between the spin vector of the pulsar and the orbital angular momentum, a consequence of the second supernova (\cite[Cameron et al. submitted]{cam_J1757}).

\section*{Acknowledgements}

The Parkes Observatory is part of the Australia Telescope National Facility which is funded by the Australian Government for operation as a National Facility managed by CSIRO. The Green Bank Observatory is a facility of the National Science Foundation operated under cooperative agreement by Associated Universities, Inc. Pulsar research at the Jodrell Bank Centre for Astrophysics and the observations using the Lovell Telescope are supported by a consolidated grant from the STFC in the UK. This work is also based on observations with the 100-m telescope of the Max-Planck-Institut f{\"u}r Radioastronomie at Effelsberg. Survey processing was conducted at the Australian National Computational Infrastructure high performance computing centre at the Australian National University, in association with the ARC Centre of Excellence for All-sky Astrophysics. AC acknowledges the support of both the International Max Planck Research School for Astronomy and Astrophysics at the Universities of Bonn and Cologne, and the Bonn-Cologne Graduate School of Physics and Astronomy.

\end{document}